\documentclass[preprint,aps,floats]{revtex4}
\usepackage{epsfig}
\usepackage{graphicx}% Include figure files
\usepackage{dcolumn}% Align table columns on decimal point
\usepackage{bm}% bold math
\def\lsim{\mathrel {\vcenter {\baselineskip 0pt \kern 0pt
    \hbox{$<$} \kern 0pt \hbox{$\sim$} }}}
\def\gsim{\mathrel {\vcenter {\baselineskip 0pt \kern 0pt
    \hbox{$>$} \kern 0pt \hbox{$\sim$} }}}

\begin{document}

\title{SUSY R-parity violating contributions to the width differences
for $D-\bar{D}$ and $B_{d,s}-\bar B_{d,s}$ systems}

\author{Shao-Long Chen}
\email{shaolong@phys.ntu.edu.tw}
\author{Xiao-Gang He}
\email{hexg@phys.ntu.edu.tw}
\author{A.~Hovhannisyan\footnote{On leave from Yerevan Physics Institute, Yerevan, Armenia.}}
\email{artyom@phys.ntu.edu.tw}
\author{Ho-Chin Tsai}
\email{hctsai@phys.ntu.edu.tw}
\affiliation{
Department of Physics and Center for Theoretical Sciences, National
Taiwan University, Taipei, Taiwan}

\date{\today}

%abstract=================================================================
\begin{abstract}
We study R-parity violating contributions to the mixing parameter
$y$ for $D^0 -\bar D^0$ and $B^0_{d,s} - \bar B^0_{d,s}$ systems.
We first obtain general expressions for new physics contributions
to $y$ from effective four fermion operators. We then use them to
study R-parity contributions. We find that R-parity violating
contributions to $D^0 - \bar D^0$ mixing, and $B_{d}^0 - \bar
B_{d}^0$ to be small. There may be sizable contribution to $B_s^0
-\bar B_s^0$ mixing. We also obtain some interesting bounds on
R-parity violating parameters using known Standard Model
predictions and experimental data.

\end{abstract}

\maketitle
\section{Introduction}
Mixing between a neutral meson with specific flavor and its
anti-meson provides powerful test for the Standard Model (SM) and
new physics (NP) beyond.  Mixing has been observed in
\cite{Yao:2006} $K^0-\bar K^0$, $B^0_d -\bar B^0_d$, and $B_s^0-\bar
B^0_s$. Evidence for $D^0$ mixing has been recently reported by the
BaBar \cite{Aubert:2007} and Belle \cite{Abe:2007, Staric:2007}
collaborations.

Two parameters, $x = \Delta M/ \Gamma$ and $y = \Delta
\Gamma/2\Gamma$, are often used to describe the mixing between a
meson and its anti-meson. Here $\Gamma$ is the life-time of the
meson. $\Delta M = m_2 - m_1$, $\Delta \Gamma = \Gamma_2 - \Gamma_1$
with ``1'' and ``2'' indicating the CP odd and CP even states,
respectively, in the limit of CP conservation. $\Delta M$ and
$\Delta \Gamma$ are related to the mixing matrix elements $M_{12}$
and $\Gamma_{12}$ in the Hamiltonian by $\Delta M - i \Delta
\Gamma/2 = 2\sqrt{(M_{12}-i\Gamma_{12}/2)(M_{12}^* - i
\Gamma_{12}^*/2)}$.

If a new particle has flavor changing neutral current (FCNC)
interaction, a non-zero contribution to $M_{12}$ can be easily
generated by exchanging this new particle in the intermediate state,
tree or loop. The parameter $\Gamma_{12}$ must come from the
absorptive part which requires the intermediate states be light
degrees of freedom to whom the meson can decay into. This fact
severely constrains the contributions to $\Gamma_{12}$ from NP. Due
to this reason there is less theoretical work on new physics
contributions to $\Gamma_{12}$ than that for $M_{12}$. In this work,
we study the $\Gamma_{12}$ parameter in the present of NP, taking
SUSY R-parity violating (RPV) interaction as an explicit example.

There are three types of $R$-Parity violating (RPV) terms
\cite{rbreaking}:
\begin{eqnarray}
&&{\lambda_{ijk}\over 2} L_L^iL_L^j  E_R^{ck}\;,\;\;\lambda'_{ijk}
L_L^iQ_L^j D_R^{ck}\;, \;\;{\lambda''_{ijk}\over 2} U_R^{ci}
D_R^{cj} D_R^{ck} \; ,
\end{eqnarray}
where $i,j$ and $k$ are the generation indices: $L_L^{}, Q_L^{},
E_R^{}, D_R^{}$ and $U_R^{}$ are the chiral superfields which
transform under the SM gauge group $SU(3)_C\times SU(2)_L\times
U(1)_Y$ as $L_L: (1,2,-1)$, $E_R: (1, 1, -2)$, $Q_L: (3,2,1/3)$,
$U_R:(3,1,4/3)$ and $D_R: (3,1, -2/3)$. The charge conjugated field
$\psi^c_R$ is defined as  $\psi^c_R = C({\bar \psi_R})^T$. We will
consider each of these R-parity contributions to $\Delta
\Gamma_{12}$ for meson mixing separately. In that case, as the term
proportional to $\lambda_{ijk}$ involves only leptons, it will not
contribute to meson mixing, since we are not considering pairs
of $\lambda_{ijk}$ and $\lambda'_{ijk}$ couplings to be non-zero at the same time. We only need to consider the last two terms up to one loop level.

At the tree level, we have the following terms relevant to us by
exchange s-fermions,
\begin{eqnarray}
{\cal L}_{eff}(\lambda') &=&
{\lambda'_{ijk}\lambda'^*_{i'j'k}\over2m^2_{\tilde d^k_R}}\bar
e^{i'}_L\gamma^\mu e^i_L \bar u^{j'}_L\gamma_\mu u^j_L -{\lambda'
_{ijk}\lambda'^*_{ij'k'}\over 2m^2_{\tilde e^i_L}} \bar
u^{j'}_{L\beta} \gamma^\mu u^j_{L\alpha} \bar d^k_{R\alpha}
\gamma_\mu d^{k'}_{R\beta}\nonumber\\
&-&{\lambda' _{ijk}\lambda'^*_{ij'k'}\over 2m^2_{\tilde \nu^i_L}}
\bar d^{j'}_{L\alpha} \gamma^\mu d^j_{L\beta} \bar d^k_{R\beta}
\gamma_\mu d^{k'}_{R\alpha}+
{\lambda'_{ijk}\lambda'^*_{i'j'k}\over2m^2_{\tilde d^k_R}} \bar
\nu^{i'}_L\gamma^\mu \nu^i_L \bar d^{j'}_L\gamma_\mu d^j_L
\nonumber\\
& -&{\lambda' _{ijk}\lambda'^*_{i'jk'}\over 2m^2_{\tilde d^j_L}}
\bar \nu^{i'}_L \gamma^\mu \nu^i_L \bar d^k_R \gamma_\mu d^{k'}_R
-{\lambda' _{ijk}\lambda'^*_{i'jk'}\over 2m^2_{\tilde u^j_L}} \bar
e^{i'}_L \gamma^\mu e^i_L \bar d^k_R \gamma_\mu d^{k'}_R
\;\nonumber\\
{\cal L}_{eff}(\lambda'') &=& {\lambda''_{ijk}\lambda''^*_{i'jk'}
\over 2m^2_{\tilde d_R^j}} (\bar u^i_{R\alpha} \gamma^\mu
u^{i'}_{R\alpha} \bar d^k_{R\beta} \gamma_\mu d^{k'}_{R\beta} - \bar
u^i_{R\alpha} \gamma^\mu u^{i'}_{R\beta}
\bar d^k_{R\beta} \gamma_\mu d^{k'}_{R\alpha})\nonumber\\
&+& {\lambda''_{ijk}\lambda''^*_{ij'k'}\over 4m^2_{\tilde u_R^i}}
(\bar d^j_{R\alpha} \gamma^\mu d^{j'}_{R\alpha} \bar d^k_{R\beta}
\gamma_\mu d^{k'}_{R\beta} - \bar d^j_{R\alpha} \gamma^\mu
d^{j'}_{R\beta} \bar d^k_{R\beta} \gamma_\mu d^{k'}_{R\alpha})\;.
\end{eqnarray}
The first two terms in ${\cal{L}}_{eff}(\lambda^{'('')})$ contribute
to $\Gamma_{12}$ for $D^0- \bar D^0$ mixing. Except the first term
in ${\cal {L}}_{eff}(\lambda')$, all terms contribute to $B^0_{d,s}
- \bar B^0_{d,s}$ mixing.

It is clear that from the above Lagrangian at the tree level,
non-zero $M_{12}$ can be generated. Constraints have been obtained
using $\Delta M$ for various meson mixing. However, in order to
generate a non-zero $\Gamma_{12}$ additional loop corrections are
needed from the above four fermion interactions.

There are short and long distance contributions to $y$ or
$\Gamma_{12}$. The calculations for long distance contributions are
very difficult to handle due to our poor understanding of QCD at low
energies. It is expected that long distance contributions become
less and less important when energy scale becomes higher and higher,
and perturbative short distance contributions will become the
dominant one. We therefore will restrict ourselves to mesons
containing a heavy $c$ or $b$ quark and to study the short distance
contributions $\Gamma_{12}$ for $D^0 -\bar D^0$ and $B^0_{d,s} -
\bar B^0_{d,s}$ systems.

For $B_{d,s}$ mesons, in the SM the short distance contributions are
expected to be the dominant ones. The prediction for
$\Delta{\Gamma}$ for $B^0_s-\bar B^0_s$ is \cite{Lenz:2006}
\begin{eqnarray}
\Delta{\Gamma_s}=( 0.106\pm 0.032 ) \;\mbox{ps}^{-1}.
\end{eqnarray}
which gives $y_{SM}=0.078\pm 0.025$. The D{\O} experiment has
measured this width difference\cite{Abazov:2006} (see also
\cite{Acosta:2005}). Allowing the non-zero CP violation in mixing
they obtained, $\Delta \Gamma_s = (0.17\pm 0.09_{stat}\pm0.03_{syst}
)$ ps$^{-1}$ ($y=0.125\pm 0.066_{stat}\pm 0.022_{syst}$), and in the
CP conserving limit, $\Delta{\Gamma_s}=(0.12\pm 0.08_{stat-0.04
syst}^{\hspace{0.5cm}+0.03})$ ps$^{-1}$ ($y=0.088\pm
0.059_{stat-0.030 syst}^{\hspace{0.5cm}+0.022}$). Within error bars,
SM agrees with data. However, it is interesting to see if NP
contributions can appear at the SM level.

For $B_d^0-\bar B^0_d$ system the width difference in SM is known to
be small \cite{Lenz:2006} $\Delta{\Gamma_d}=(26.7\pm
0.08)\times{10^{-4}}$ ps$^{-1}$, corresponding to $y_{SM}=(2.058\pm
0.006)\times 10^{-3}$. There is no experimental data for the width
difference yet. It is interesting to see whether the width
difference can be much larger when going beyond the SM.

For the $D^0-\bar{D^0}$ mixing in the SM as was shown in
\cite{Falk:2001} $x$ and $y$ are generated only at the second order
in $SU(3)$ breaking, $ x,y\sim \sin^2{\theta_C}\times [SU(3)\,
\mbox{breaking}]^2$. Most of the studies give $x, y < 10^{-3}$,
although large values are not excluded~\cite{Donoghue:1985}.

Recently, the parameter $y$ for $D^0-\bar{D^0}$ mixing has been
measured. BaBar, assuming no CP violation in mixing, has analyzed
the doubly Cabibbo suppressed (DCS) $D^0\to K^+\pi^-$ mode
\cite{Aubert:2007}, while Belle has studied singly Cabibbo
suppressed (SCS) $D^0\to K^+K^-,\pi^+\pi^-$ decays \cite{Abe:2007}.
From these results the authors in \cite{Ciuchini:2007} have fitted
the mixing parameters and get the following result for $y$ with 68\%
and 95\% probability correspondingly
\begin{eqnarray}
y&=&(6.1\pm 1.9)\times 10^{-3},\;\; y\in  [0.0023, 0.0102].
\end{eqnarray}

In this work, we find that R-parity violating contribution to the
parameter $y$ is small for $D^0 -\bar D^0$ system, less than
 a few times $10^{-4}$. The RPV contribution for $B_d^0 - \bar B_d^0$ system
can be larger than the SM prediction. For $B_s^0 -\bar B^0_s$
system, the contribution to $y$ can be as large as the SM
contribution.

In the following sections, we provide the detailed calculations.

\section{General expression for $\Gamma_{12}$}

Before going into specific RPV model calculations, we summarize some
general results for short distance NP contribution to $\Gamma_{12}$
from four quark operators generated by SM and NP. The calculation is
straightforward. Starting from tree level four quark interactions,
one needs to obtain the absorptive part for Fig.1. Let us take $D^0
-\bar D^0$ mixing for illustration. For the cases considered here,
we can write the $\Delta C=-1$ interaction Lagrangian as
\begin{eqnarray}
&& {\cal L}^{\Delta C=-1}= \sum_{q,q'}\left\{ {\bf D}_{qq'}
\left[{\cal C}_1(\mu) Q_1 + {\cal C}_2 (\mu) Q_2 \right] +{\bf
D'}_{qq'} \left[{\cal C'}_1(\mu) Q_3 + {\cal C'}_2 (\mu) Q_4 \right]
\right\}\;,\label{opt}
\\
&& Q_1 = \overline{u}_i \Gamma_1 q'_j ~\overline{q}_j \Gamma_2 c_i \
, \;\;Q_2 = \overline{u}_i \Gamma_1 q'_i~\overline{q}_j \Gamma_2
c_j\;, \;\; Q_3 = \overline{u}_i \Gamma_3 q'_j ~\overline{q}_j
\Gamma_4 c_i \ , \;\;Q_4 = \overline{u}_i \Gamma_3
q'_i~\overline{q}_j \Gamma_4 c_j\;. \nonumber
\end{eqnarray}
In the above we have omitted possible Lorentz indices for $\Gamma_i$
which are contracted. The specific form of $\Gamma_i$ depends on the
nature of interaction generating the four quark operators.  The
notations here are that $\Gamma_{1,2 (3,4)}$ and $\Gamma_{3,4
(1,2)}$ should appear on the left and right four quark vertices in
Fig. 1, respectively.

\begin{figure}[th!]
\begin{center}\label{feynman}
\includegraphics[width=4.0 in]{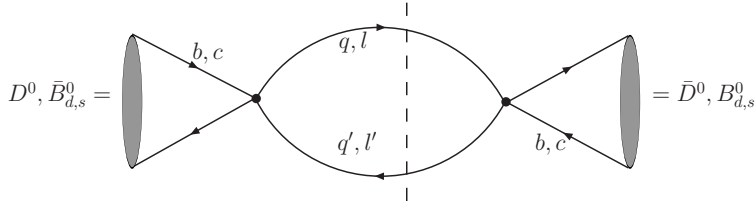}
\caption{\small The one-loop Feynman diagram for
meson mixing. The dashed line represents the cut for taking the
absorptive part.}
\end{center}
\end{figure}

Evaluating the diagram in Fig. 1, one obtains the following general
expression for $\Gamma_{12}$,
\begin{eqnarray}\label{yformula}
\Gamma_{12} = - \frac{1}{2M_{\rm D} } \ \sum_{q,q'} {\bf D}_{qq'}
{\bf D'}_{q'q} \left(K_1 \delta_{ik}\delta_{j\ell} + K_2
\delta_{i\ell}\delta_{jk} \right) \sum_{\alpha=1}^5 \; I_\alpha
(x,x') \; \langle \overline{D}^0| \; {\mathcal O}_\alpha^{ijk\ell}
\; | D^0 \rangle, \ \
\end{eqnarray}
where
\begin{eqnarray}
K_1= \left({\mathcal C}_1 {\mathcal C'}_1 N_c + \left({\mathcal C}_1
{\mathcal C'}_2 + {\mathcal C'}_1 {\cal C}_2 \right)\right), \ K_2 =
{\mathcal C}_2 {\mathcal C'}_2 .\label{lepton}
\end{eqnarray}

The operators are defined as
\begin{eqnarray}
&&{\cal O}_1^{ijk\ell} \ = \ \overline{u}_k \Gamma_3 \gamma_\nu
\Gamma_2 c_j ~ \overline{u}_\ell \Gamma_1 \gamma^\nu \Gamma_4 c_i
,\;\; {\cal O}_2^{ijk\ell} \ = \ \overline{u}_k \Gamma_3
\hspace{-0.14cm} \not p_c \Gamma_2 c_j ~ \overline{u}_\ell \Gamma_1
\not p_c \Gamma_4 c_i ,
\nonumber \\
&&{\cal O}_3^{ijk\ell} \ = \ \overline{u}_k \Gamma_3 \Gamma_2 c_j ~
\overline{u}_\ell \Gamma_1 \hspace{-0.14cm} \not p_c \Gamma_4 c_i ,
\;\;\;\;\; {\cal O}_4^{ijk\ell} \ = \ \overline{u}_k \Gamma_3
\hspace{-0.14cm} \not p_c \Gamma_2 c_j ~
\overline{u}_\ell \Gamma_1 \Gamma_4 c_i , \nonumber\\
&&{\cal O}_5^{ijk\ell} \ = \ \overline{u}_k \Gamma_3 \Gamma_2 c_j ~
\overline{u}_\ell \Gamma_1 \Gamma_4 c_i  , \nonumber
\end{eqnarray}
and the coefficients $I_\alpha(x,x')$ are given by
\begin{eqnarray}
&&I_1(x,x')=-\frac{k^* m_c}{48\pi}\left[ 1-
2\left(x+x'\right)+\left(x-x'\right)^2 \right] ,\nonumber\\
&&I_2(x,x')=-\frac{k^*}{24\pi m_c} \left[
1+\left(x+x'\right)-2\left(x-x'\right)^2 \right]
\nonumber \\
&&I_3(x,x')=\frac{k^*}{8\pi}  \sqrt{x} \left(1+x'-x\right)
,\;\;I_4(x,x')=-\frac{k^*}{8\pi} \sqrt{x'}
\left(1-x'+x\right),\nonumber\\
&& I_5(x,x')=\frac{k^* m_c}{4\pi} \sqrt{x x'} \ \ ,\label{Is}
\end{eqnarray}
where $k^* \equiv (m_c/2)[1-2 (x + x') + (x - x')^2]^{1/2}$ with
$x\equiv m_q^2/m_c^2$ and $x'\equiv m_{q'}^2/m_c^2$. Replacing
$D'_{qq'}$ with the SM couplings and $\Gamma_{3}=
\gamma^\mu(1-\gamma_5)/2$ and $\Gamma_4 = \gamma_\mu
(1-\gamma_5)/2$, we obtain the formula presented in
\cite{Golowich:2006,Golowich:2007ka} for SM and NP interference
contribution.  Note that when considering the contributions from the
same operator, one should take $D'_{qq'} = D_{qq'}$, $\Gamma_{1,2} =
\Gamma_{3,4}$ and the expression for $\Gamma_{12}$ should be divided
by 2.

Using the above formula, one can easily work out the expressions
contributing from the SM (taking SM operators for $Q_{1,2,3,4}$),
the interference between the SM and NP (SM - NP) (taking $Q_{1,2}$
from SM (NP) and $Q_{3,4}$ from NP (SM)), and purely NP (NP - NP)
(taking $Q_{1,2,3,4}$ from NP). New physics effects can show up in
the later two cases. We will concentrate on these contributions.

We comment that the fermions in the loop are not necessary to be
quarks. They can be leptons too. If one identifies $q$ and $q'$ to
be leptons, the correct result can be obtained by setting $N_c = 1$
and $C_2^{(')}=0$ in Eq.~(\ref{lepton}). One can easily generalize
the above formula for $B^0_{d,s} - \bar B^0_{d,s}$ mixing cases with
appropriate replacement of quark fields and couplings.

\section{RPV contributions to $\Gamma_{12}$ for $D^0 - \bar D^0 $ mixing}

In this section we give expressions for contributions to
$\Gamma_{12}$ with RPV interactions.

 Contributions from $\lambda'$
interaction to $\Gamma_{12}$ are given by
\begin{eqnarray}\label{drpv1}
&&{\Gamma_{12}}_{(SM-RPV)}=\frac{\sqrt{2}G_F
\lambda^\prime_{i22}\lambda^{\prime*}_{i12}V_{us}V^*_{cs}} {8\pi m_D
m^2_{\tilde e^i_L}}x_sm_c^2 (C_1+C_2)\langle Q \rangle
\;,\nonumber\\
&&{\Gamma_{12}}_{(RPV-RPV,
l)}=\frac{\lambda'_{i2k}\lambda'^{*}_{j1k}
\lambda^\prime_{j2k^\prime}\lambda^{\prime*}_{i1k^\prime}} {192\pi
m^2_{\tilde d^k_R}m^2_{\tilde d^{k^\prime}_R}}
\frac{m_c^2}{m_D}\left(\langle Q \rangle + \langle Q_s
\rangle\right)\;,\\
&& {\Gamma_{12}}_{(RPV-RPV,d)} =
\frac{\lambda'_{i2j'}\lambda'^{*}_{i1j}
\lambda^\prime_{i'2j}\lambda^{\prime*}_{i'1j'}} {192\pi m^2_{\tilde
e^i_L}m^2_{\tilde e^{i'}_L}}
\frac{m_c^2}{m_D}\left(\frac{1}{2}\langle Q \rangle - \langle
Q_s\rangle\right),\nonumber
\end{eqnarray}
where
\begin{eqnarray}
\nonumber \langle Q \rangle &=& \langle \bar D^0|\bar u_\alpha
\gamma^\mu P_Lc_\alpha \bar u_\beta \gamma_\mu P_L c_\beta
|D^0\rangle \;,\;\; \langle Q_s \rangle = \langle \bar D^0|\bar
u_\alpha P_R c_\alpha \bar u_\beta P_R c_\beta |D^0\rangle  \;,
\end{eqnarray}
The first equation in Eq.~(\ref{drpv1}) is the leading order result in $x_s$.
Depending on the internal lepton exchanges, in the expression for
${\Gamma_{12}}_{(RPV-RPV,l)}$ the indices $i, j$ take 1 and 2
indicating which charged leptons are in the loop. In principle, one
can also have an electron and a tauon in the loop. However, the
tauon mass is close to the D meson mass, the contribution is
suppressed by phase space. We will neglect this contribution. In the
expression for ${\Gamma_{12}}_{(RPV-RPV,q)}$, $j,j'$ take 1 and 2
indicating which of the down quarks are in the loop.

${\Gamma_{12}}_{(SM-RPV)}$ comes from SM interaction with the second
term, ${\Gamma_{12}}_{(RPV-RPV,l)}$ comes from the first term, and
${\Gamma_{12}}_{(RPV-RPV, q)}$ comes from the second term, in ${\cal
{L}}_{eff}(\lambda')$, respectively. Note that the SM-RPV
contribution is proportional to the internal quark masses and the
dominant one comes from $s\bar{s}$ in the loop. This is due to the
chiral structure of $\Gamma_i$ which allow only $O_5^{ijkl}$ to
contribute and therefore proportional to the function $I_5(x,x')$.
If $x$ or $x'$ takes the down quark mass $I_5(x,x')$ is negligibly
small.

In obtaining the expression for ${\Gamma_{12}}_{(SM-RPV)}$, we have
used the SM $\Delta C = -1$ Lagrangian,
\begin{eqnarray}
{\cal {L}}_{SM} = - {4G_F\over \sqrt{2}}
V^*_{qc}V_{q'u}\left[C_1(m_c) Q_1 + C_2(m_c) Q_2\right],
\end{eqnarray}
with $\Gamma_1$ and $\Gamma_2$ in Eq.(\ref{opt}) to be $\gamma_\mu
(1-\gamma_5)/2$ and $\gamma^\mu (1-\gamma_5)/2$, respectively.

The contributions from $\lambda''$ interaction come from the first
term in ${\cal {L}}_{eff}(\lambda'')$ and are given by
\begin{eqnarray}
&&{\Gamma_{12}}_{(SM-RPV)} =-\frac{\sqrt{2}G_F
\lambda''_{1j2}\lambda''^{*}_{2j2}V_{us}V^*_{cs}} {8\pi m_D
m^2_{\tilde d^j_R}}x_s m_c^2 [(2C_1+C_2) \langle
Q' \rangle -C_2 \langle \tilde Q' \rangle ]\;,\nonumber\\
&&{\Gamma_{12}}_{(RPV-RPV)}=
\frac{\lambda''_{1ji}\lambda''^{*}_{2ji'}
\lambda''_{1j'i'}\lambda''^*_{2j'i}} {192\pi m^2_{\tilde
d^j_R}m^2_{\tilde d^{j'}_R}}
\frac{m_c^2}{m_D}\left(\frac{3}{2}\langle Q
\rangle\right)\;.\label{squark1}
\end{eqnarray}
where
\begin{eqnarray}
\nonumber \langle Q^\prime \rangle &=& \langle \bar D^0|\bar
u_\alpha \gamma^\mu P_Lc_\alpha \bar u_\beta \gamma_\mu P_R c_\beta
|D^0\rangle \;,\;\; \langle \tilde Q' \rangle = \langle \bar
D^0|\bar u_\alpha \gamma^\mu P_Lc_\beta \bar u_\beta \gamma_\mu P_R
c_\alpha |D^0\rangle \;.
\end{eqnarray}
In first equation of Eq.~(\ref{squark1}), as in the first equation of
Eq.~(\ref{drpv1}), we only kept the leading order in $x_s$. The SM-RPV
contribution is dominated by $s\bar{s}$ pair in the loop for the
same reason as that for the $\lambda'$ case for SM-RPV contribution
explained earlier.

Here we should mention that recently in Ref.~\cite{Golowich:2006}
the authors have considered RPV with slepton and squark exchanges
for SM-NP contributions. Our predictions in the first equations in
Eqs.~(\ref{drpv1}) and (\ref{squark1}), for the same measurable, do not agree
with their Eqs.~(16) and (24), respectively.

\section{RPV contributions to $\Gamma_{12}$ for $B^0_{d,s}-\bar B^0_{d,s}$ mixing}

In this case all terms except the first term in ${\cal
{L}}_{eff}(\lambda')$ contribute to $\Gamma_{12}$.

\subsection{The $\lambda'$ contribution}

The expressions for $\Gamma_{12}$ from various contributions are
given by
\begin{eqnarray}
&& {\Gamma_{12}}_{(SM-RPV)}=\frac{\sqrt{2}G_F
m_b^2\lambda_{q'qk}}{48\pi m_B m^2_{\tilde {e}^i_L} } \left\{
(2C_1(m_b)-C_2(m_b))\langle Q'\rangle +(2C_2(m_b)-C_1(m_b))\langle
{\tilde {Q'}}\rangle \right\}\;,\nonumber\\
&& {\Gamma_{12}}_{(RPV-RPV,\nu)}=\frac{m_b^2}{192\pi m_B}\left\{
\frac{\lambda'_{j3i'}\lambda'^*_{j'ki'}}{m^2_{\tilde{d}^{i'}_R}}
\frac{\lambda'_{j'3i}\lambda'^*_{jki}}{m^2_{\tilde{d}^i_R}}
\left(\langle Q\rangle + \langle Q_s \rangle \right)\right. \nonumber\\
&&\hspace{2.5cm}\;\;+
\left.\frac{\lambda'_{jik}\lambda'^*_{j'i3}}{m^2_{\tilde{d}^i_L}}
\frac{\lambda'_{j'i'k}\lambda'^*_{ji'3}}{m^2_{\tilde{d}^{i'}_L}}
\left(\langle Q\rangle + \langle Q_s \rangle \right) -2
\frac{\lambda'_{j3i'}\lambda'^*_{j'ki'}}{m^2_{\tilde{d}^{i'}_R}}
\frac{\lambda'_{j'ik}\lambda'^*_{ji3}}{m^2_{\tilde{d}^i_L}}
\left(\langle Q'\rangle -\frac{1}{2} \langle \tilde{Q}' \rangle
\right)
\right\} \;,\nonumber\\
&&{\Gamma_{12}}_{(RPV-RPV,l)}= \frac{m_b^2}{192\pi m_B}
\frac{\lambda'_{jik}\lambda'^*_{j'i3}}{m^2_{\tilde{u}^i_L}}
\frac{\lambda'_{j'i'k}\lambda'^*_{ji'3}}{m^2_{\tilde{u}^{i'}_L}}
\left( \langle Q_s\rangle + \langle Q\rangle \right)\;,\\
&&{\Gamma_{12}}_{(RPV-RPV,u)}= \frac{m_b^2}{192\pi m_B}
\frac{\lambda'_{ijk}\lambda'^{*}_{ij'3}}{m^2_{\tilde e^i_L}}
\frac{\lambda'_{i'j'k}\lambda'^{*}_{i'j3}}{m^2_{\tilde e^{j'}_L}}
\left(\frac{1}{2}\langle Q \rangle - \langle  {Q_s}
\rangle\right)\;,\nonumber\\
&&{\Gamma_{12}}_{(RPV-RPV,d)} = \frac{m_b^2}{192\pi m_B}
\frac{1}{m^2_{\tilde {\nu}^i_L} m^2_{\tilde {\nu}^{i'}_L}} \left\{
18 \lambda'_{ijj'}\lambda'^*_{ik3}\lambda'_{i'3k}\lambda'^*_{i'jj'}
\langle \tilde {Q'}\rangle \right. \nonumber\\
&&\hspace{2.5cm}\;\;+
\left.\left(\lambda'_{i3j'}\lambda'^*_{ikj}\lambda'_{i'3j}\lambda'^*_{i'kj'}+
\lambda'_{ijk}\lambda'^*_{ij'3}\lambda'_{i'j'k}\lambda'^*_{i'j3}
\right) \left( \frac{1}{2} \langle Q \rangle - \langle Q_s \rangle
\right)  \right\}\;,\nonumber
\end{eqnarray}
where $j,j'$ take the values 1 and 2.
\begin{eqnarray*}
\lambda_{cck}&=& V_{cq^k}^*V_{cb}\lambda_{i2k}'\lambda_{i23}'^*
\;,\;\;
\lambda_{cuk}= V_{cq^k}^*V_{ub}\lambda_{i1k}'\lambda_{i23}'^* \;,\\
\lambda_{uck}&=& V_{uq^k}^*V_{cb}\lambda_{i2k}'\lambda_{i13}'^*
\;,\;\; \lambda_{uuk}=
V_{uq^k}^*V_{ub}\lambda_{i1k}'\lambda_{i13}'^* \;.
\end{eqnarray*}
and
\begin{eqnarray*}
\langle Q'\rangle &=& \langle B^0_{q^k}|\bar q^k_\alpha \gamma_\mu
P_L b_\alpha
~\bar q^k_\beta \gamma_\mu P_R b_\beta  |\bar B^0_{q^k}\rangle \;,\\
\langle \tilde {Q'}\rangle &=& \langle B^0_{q^k}|\bar q^k_\alpha
\gamma_\mu P_L b_\beta ~\bar q^k_\beta \gamma_\mu P_R b_\alpha |\bar
B^0_{q^k}\rangle \;.
\end{eqnarray*}
Here for $B_d$ and $B_s$ systems, $k$ takes 1 and 2, respectively.

The five different contributions to $\Gamma_{12}$ listed above come
from the second, first and sixth, fourth and fifth, second and third
terms in ${\cal {L}}_{eff}(\lambda')$, respectively.

\subsection{The $\lambda ''$ Contribution}

In this case we have
\begin{eqnarray}
&&{\Gamma_{12}}_{(SM-RPV)} = -\frac{\sqrt{2}G_F x_c\sqrt{1-4x_c}
m_b^2V^*_{cq^k}V_{cb}\lambda_{2ki}''\lambda_{23i}''^*} {8\pi
m_B m^2_{\tilde {d}^i_R} } \nonumber\\
&&\hspace{2.4cm}\times\left\{ (2C_1(m_b)+C_2(m_b))\langle Q'\rangle
- C_2(m_b)\langle {\tilde
{Q'}}\rangle \right\},\nonumber\\
&&{\Gamma_{12}}_{(RPV-RPV, u)}=\frac{m_b^2}{192\pi m_B}
\frac{\lambda''_{ijk}\lambda''^{*}_{i'j3}}{m^2_{\tilde d^j_R}}
\frac{\lambda''_{i'j'k}\lambda''^*_{ij'3}}{m^2_{\tilde d^{j'}_R}}
\left(\frac{3}{2}\langle Q \rangle\right)\;,\\
&&{\Gamma_{12}}_{(RPV-RPV,d)}=\frac{m_b^2} {48\pi m_B}
\frac{\lambda''_{j'ki}\lambda''^{*}_{j'3i'}}{m^2_{\tilde u^{j'}_R}}
\frac{\lambda''_{jki'}\lambda''^{*}_{j3i}}{m^2_{\tilde u^j_R}}
\left(\frac{3}{2}\langle Q \rangle\right)\;,\nonumber
\end{eqnarray}
where $i,i'$ take the values 1 and 2. The first two terms are due to
the first term in ${\cal {L}}_{eff}(\lambda'')$, and the last term
is due to the second term in ${\cal {L}}_{eff}(\lambda'')$.

The SM-RPV interference is dominated by $c\bar c$ exchange in the
loop for the same reasons as that for the $s\bar s$ dominance for
SM-RPV $D^0-\bar D^0$ mixing.

\section{Numerical analysis}

In this section, we carry out numerical analysis for RPV
contributions to mixing parameter $y$ for $D^0-\bar D^0$ and
$B^0_{d,s} - \bar B^0_{d,s}$ systems.

In general RPV contribution to $\Gamma_{12}$ has CP violating phases
associated with the new couplings. The relation of $\Gamma_{12}$ and
$y$ is not trivial. If CP violating effects can be neglected which
is the case for the SM, they have a simple relation
\begin{eqnarray}
y\equiv\frac{\Gamma_{12}}{\Gamma}.
\end{eqnarray}

In our numerical analysis, we assume CP conservation for easy
comparison with data and other constraints obtained in the
literature.

To compare with data, one needs to evaluate various hadronic matrix
elements in the expressions for $\Gamma_{12}$.  We write them in the
following form
\begin{eqnarray}
\nonumber \langle Q \rangle &=& \frac{2}{3}f_P^2m_P^2 B_{Q}\;, \;\;
\langle Q_s \rangle = -\frac{5}{12}f_P^2m_P^2 B_{Q_s}\;,
\\
\langle Q' \rangle &=& -\frac{5}{6}f_P^2m_P^2 B_{Q'}\;, \;\; \langle
\tilde Q' \rangle = -\frac{7}{6}f_P^2m_P^2 B_{\tilde Q'}\;,
\end{eqnarray}
where $B_{Q}$ factors are the so called bag parameters
\cite{Beneke:1996}. This way of parameterizing the matrix elements
was inspired by vacuum saturation approximation. In the vacuum
saturation approximation, they are all equal to one, which we will
use in our estimate.

\begin{table}[thb!]
\begin{center}
{\small
\begin{tabular}{ccccc}
\hline\hline ${\mu} = 1.3$~GeV  &  ${\mu} = 4.8$~GeV
  &  Masses (GeV)   &  Decay cons. (GeV)
  &  Widths (GeV)    \vspace*{-0cm} \\
 \cite{Lenz:2006, Beneke:1996}  &  \cite{Lenz:2006, Beneke:1996}
  &  \cite{Yao:2006}   &  \cite{Okamoto:2005zg}
  & \cite{Yao:2006}
\\ \hline
  $x_s=0.006$  \ & $x_c=0.0841$ & $M_{D}=1.8645$  &  $f_D=0.201$ &  $\Gamma_{D}=1.6\times 10^{-12}$ \\
  $C_1=-0.411$ \ & $C_1=-0.272$ & $M_{B_d}=5.279$ & $f_B=0.216$  &  $\Gamma_{B_d}=4.27\times
10^{-13}$\\
  $C_2=1.208$  \ & $C_2=1.120$  & $M_{B_s}=5.368$ & $f_{B_s}=0.260$ & $\Gamma_{B_s}=4.46\times 10^{-13}$ \\
  \hline\hline
\end{tabular}
\caption{The central values of input parameters and coefficients.}}
\end{center}
\end{table}

In the Table I we list the parameters and coefficients appearing in
the equations above. The input CKM elements are \cite{Charles:2004}
\begin{eqnarray}
\lambda\equiv |V_{us}|=0.2248 ,\;\;\
A\lambda^2\equiv|V_{cb}|=41.5\times{10^{-3}}.
\end{eqnarray}
The charm quark mass also comes into the calculations. In our
numerical analysis we identify $m_c$ and $m_b$ with \emph{pole}
masses. Numerically we use \cite{Beneke:1996} $m_{c,pole}/m_{b,pole}
= \sqrt{x_c} =0.29$, which is based on the mass difference
$m_{c,pole}-m_{b,pole}=3.4$ GeV and $m_{b,pole}=4.8$ GeV.

To give some understanding of RPV contributions, in the following
analysis we take the central values for the input parameters.

\subsection{$D^0-\bar D^0$ mixing}
Taking all $\lambda'$ and $\lambda''$ to be real, and inserting
known values for the parameters involved, for $\lambda'$
contributions, we have
\begin{eqnarray}\label{eq:D0}
&&y_{(SM-RPV)}=0.0037\times\lambda^\prime_{i22}\lambda^{\prime*}_{i12}\frac{(100GeV)^2}
{m^2_{\tilde e^i_L}}
\;,\nonumber\\
&&y_{(RPV-RPV, l)}=0.3298\times\lambda'_{i2k}\lambda'^{*}_{j1k}
\lambda^\prime_{j2k^\prime}\lambda^{\prime*}_{i1k^\prime}
\frac{(100GeV)^4}{m^2_{\tilde d^k_R}m^2_{\tilde d^{k^\prime}_R}}\;,\\
&& y_{(RPV-RPV,q)} = 0.9893\times\lambda'_{i2j'}\lambda'^{*}_{i1j}
\lambda^\prime_{i'2j}\lambda^{\prime*}_{i'1j'}
\frac{(100GeV)^4}{m^2_{\tilde e^i_L}m^2_{\tilde e^{i'}_L}}.\nonumber
\end{eqnarray}
The $\lambda''$ contributions are given by
\begin{eqnarray}
&&y_{(SM-RPV)} =
-0.0077\times\lambda''_{1j2}\lambda''^{*}_{2j2}\frac{(100GeV)^2}
{m^2_{\tilde d^j_R}}\;,\nonumber\\
&&y_{(RPV-RPV)}= 1.3191\times\lambda''_{1ji}\lambda''^{*}_{2ji'}
\lambda''_{1j'i'}\lambda''^*_{2j'i}\frac{(100GeV)^4} {m^2_{\tilde
d^j_R}m^2_{\tilde d^{j'}_R}}\;.
\end{eqnarray}

There are constraints on the RPV parameters from various other
processes\cite{Barbier:2004ez,Chemtob:2004,Allanach:1999ic}. Taking
these constraints into account, we list in Table II the
corresponding values for the mixing parameter $y$.

\begin{table}[thb!]\label{tab:D0}
\begin{center}
\begin{tabular}{ccccccccr}
\hline\hline
 \ RPV parameters  \ & \ Bounds [Processes] \ & \ Estimate  \
\\ \hline
  $|\lambda^\prime_{i22}\lambda^{\prime*}_{i12}|$ \ & 0.07 (see text)
  & $y_{(SM-RPV)}\simeq 2.6\times{10^{-4}}$   \\
  $|\lambda'_{i2k}\lambda'^{*}_{j1k}|$ \ & $ 5.28\times 10^{-6}$ [$K^+\to\pi^+\nu\bar{\nu}$]
  & $y_{(RPV-RPV, l)}\simeq 9.2\times 10^{-12}$  \\
  $|\lambda'_{i2j}\lambda'^{*}_{i'1j}|_{j=1,2}$ \ & $5.28\times 10^{-6}$ [$K^+\to\pi^+\nu\bar{\nu}$]
  & $y_{(RPV-RPV,q)}\simeq 2.5\times10^{-11}$  \\
  $|\lambda''_{132}\lambda''^{*}_{232}|$ \ & $3.1\times 10^{-3}$ [$D\bar{D}$]
  & $y_{(SM-RPV)}\simeq 2.4\times 10^{-5}$  \\
  $|\lambda''_{132}\lambda''^{*}_{232}|$ \ & $3.1\times 10^{-3}$ [$D\bar{D}$]
  & $y_{(RPV-RPV)}\simeq 1.3\times 10^{-5}$  \\
  \hline\hline
\end{tabular}
\caption{The bounds on parameters from
\cite{Barbier:2004ez,Chemtob:2004,Allanach:1999ic,Deandrea:2004ae,Carlson:1995ji}
 and corresponding values for $y$.}
\end{center}
\end{table}

For the contribution due to the $\lambda'$ terms, using the
constraint $|\lambda'_{i2k}\lambda'^{*}_{j1k}| \lesssim 5.28\times
10^{-6}$, obtained from $K^+\to
\pi^+\nu\bar\nu$~\cite{Deandrea:2004ae} and taking the same bound
for $|\lambda'_{i2j}\lambda'^{*}_{i'1j}|_{j=1,2}$ with the
assumption that there is no accidental cancellation, we find
$y_{(RPV-RPV,l)}$ and $y_{(RPV-RPV,q)}$ to be tiny $\lesssim
10^{-11}$. For the interference term due to the SM with $\lambda'$
term, we have not found direct constraint on the appropriate
combination of the $\lambda'$ terms. We therefore use individual
constraints from \cite{Allanach:1999ic}
$(\lambda^\prime_{122}\lambda^{\prime*}_{112},
\lambda^\prime_{222}\lambda^{\prime*}_{212},
\lambda^\prime_{322}\lambda^{\prime*}_{312})=(0.0009,0.0124,0.0572)$.
This leads to  $y_{(SM-RPV)}\simeq (3.3\times 10^{-6},4.6\times
10^{-5},2.1\times 10^{-4})$. To see the largest possible value for
$y$, we sum these three with the same sign to obtain an upper bound
$y_{(SM-RPV)} = 2.6\times 10^{-4}$. This contradicts with the result
obtained in Ref.~\cite{Golowich:2006}, where $y$ can be as large as
$\simeq -3.7\%$.

As for the contributions from $\lambda''$, from the constraint
$|\lambda''_{132}\lambda''^{*}_{232}|\lesssim 3.1\times
10^{-3}$~\cite{Carlson:1995ji} ,
$\lambda''_{112},\lambda''_{121}\lesssim 2\times 10^{-9}$ and
$\lambda''_{131} < 10^{-4}$~\cite{Allanach:1999ic}, we find that
$\lambda''$ contributions are small, less than $10^{-4}$.

We conclude that R-parity violating contributions to $y$ for $D^0 -
\bar D^0$ mixing are small.

\subsection{$B_d^0-\bar B_d^0$ mixing}
Here we present the numerical results for the $B_d^0-\bar B_d^0$
mixing. For $\lambda'$ contributions, we have
\begin{eqnarray}\label{eq:Bd}
&&y_{(SM-RPV)} = -224.5\times\lambda_{q'q1}\frac{(100GeV)^2}{m^2_{\tilde {e}^i_L}}\;,\nonumber\\
&&y_{(RPV-RPV,\nu)} = 55.1\times (100\mbox{GeV})^4\left\{
\frac{\lambda'_{j3i'}\lambda'^*_{j'1i'}}{m^2_{\tilde{d}^{i'}_R}}
\frac{\lambda'_{j'3i}\lambda'^*_{j1i}}{m^2_{\tilde{d}^i_R}}
\right. \nonumber\\
&&\hspace{2.4cm}+\left.\frac{\lambda'_{ji1}\lambda'^*_{j'i3}}{m^2_{\tilde{d}^i_L}}
\frac{\lambda'_{j'i'1}\lambda'^*_{ji'3}}{m^2_{\tilde{d}^{i'}_L}} +2
\frac{\lambda'_{j3i'}\lambda'^*_{j'1i'}}{m^2_{\tilde{d}^{i'}_R}}
\frac{\lambda'_{j'i1}\lambda'^*_{ji3}}{m^2_{\tilde{d}^i_L}}
\right\} \;,\nonumber\\
&&y_{(RPV-RPV,l)}=
55.1\times\lambda'_{ji1}\lambda'^*_{j'i3}\lambda'_{j'i'1}\lambda'^*_{ji'3}
\frac{(100GeV)^4}{m^2_{\tilde{u}^i_L}m^2_{\tilde{u}^{i'}_L}}\;,\\
&&y_{(RPV-RPV,u)} =
165.2\times\lambda'_{ij1}\lambda'^{*}_{ij'3}\lambda'_{i'j'1}\lambda'^{*}_{i'j3}\frac{(100GeV)^4}{m^2_{\tilde
e^i_L}m^2_{\tilde e^{i'}_L}}  \;,\nonumber\\
&&y_{(RPV-RPV,d)} = 165.2 \times
\left(\lambda'_{i3j'}\lambda'^*_{i1j}\lambda'_{i'3j}\lambda'^*_{i'1j'}+
\lambda'_{ij1}\lambda'^*_{ij'3}\lambda'_{i'j'1}\lambda'^*_{i'j3}\right.\\&&
\hspace{2.4cm}\left.-28\lambda'_{ijj'}\lambda'^*_{i13}\lambda'_{i'31}\lambda'^*_{i'jj'}
\right) \frac{(100GeV)^4}{m^2_{\tilde {\nu}^i_L} m^2_{\tilde
{\nu}^{i'}_L}} \;.\nonumber
\end{eqnarray}

\begin{table}[t!]
\begin{center}
\begin{tabular}{ccc}
\hline\hline
 \ RPV parameters  \ & \ Bounds [Processes] \ & \ Estimate \
\\ \hline
$|\lambda'_{i21}\lambda'^*_{i13}|$ \ & $1.2\times{10^{-5}}$\,$[B\bar
B]$ \
& $y_{(SM-RPV)_1}\simeq 1.1\times 10^{-4}$ \\

$|\lambda'_{i21}\lambda'^*_{i23}|$ \ & $5.0\times{10^{-5}}$\,$[B\bar
B]$  \
& $y_{(SM-RPV)_2}\simeq 1.0\times 10^{-4}$ \\

$|\lambda_{ij1}'^*\lambda_{i'j3}'|$,~$|\lambda_{i1j}'^*\lambda_{i'3j}'|$
& $ 1.1\times 10^{-3}\;[B^0\to X_q\nu\bar\nu]$  & $y_{(RPV-RPV,\nu)}\simeq 2.7\times 10^{-4} $ \\

$|\lambda_{ij1}'^*\lambda_{i'j3}'|$  & $1.1\times{10^{-3}}$ [$B^0\to
X_q \nu\nu$]
& $y_{(RPV-RPV,l)}\simeq 0.67\times{10^{-4}}$ \\

$|\lambda'_{ij1}\lambda'^{*}_{ij'3}|\cdot|\lambda'_{i'j'1}\lambda'^{*}_{i'j3}|$
& $6.4\times{10^{-7}}$ [$\bar BB$] &   $y_{(RPV-RPV,u)}\simeq 1.1\times{10^{-4}}$ \\

$|\lambda_{ijj'}'\lambda_{i13}'|\cdot|\lambda_{ijj'}'\lambda_{i31}'|$
   \ & $1.6\times 10^{-6}$ $\;[\bar BB]$ & $y_{(RPV-RPV,d)}\simeq 7.2\times 10^{-3}$ \\
$|\lambda_{212}''\lambda_{232}''|$  & $6\times 10^{-5}$~$[B\to\phi\pi]$ & $y_{(SM-RPV)}\simeq 2.8\times 10^{-5} $\\
  $|\lambda''_{j21}\lambda''^{*}_{j23}|$ \ & $6\times{10^{-5}}$ [$B\to \phi\pi$] &
  $y_{(RPV-RPV,u)}\simeq 0.8\times 10^{-6}$\\
  $|\lambda''_{j12}\lambda''^{*}_{j32}|$ \ & $6\times{10^{-5}}$ [$B\to \phi\pi$] &
  $y_{(RPV-RPV,d)}\simeq 3.2\times 10^{-6}$ \\
  \hline\hline
\end{tabular}
\caption{The bounds on parameters from
\cite{Barbier:2004ez,Allanach:1999ic,Chemtob:2004,Kundu:2004,Ghosh:2001mr,Bar-Shalom:2002sv}
and corresponding values for $y$.}
\end{center}
\end{table}

For $\lambda''$ contributions, we have
\begin{eqnarray}
&&y_{(SM-RPV)} =0.46\times\lambda_{212}''\lambda_{232}''^*\frac{(100GeV)^2} {m^2_{\tilde {d}^2_R} },\nonumber\\
&&y_{(RPV-RPV, u)}= 220.3\times\lambda''_{i21}\lambda''^{*}_{i'23}
\lambda''_{i'21}\lambda''^*_{i23}
\frac{(100GeV)^4}{m^2_{\tilde d^2_R} m^2_{\tilde d^{2}_R}}\;,\\
&&y_{(RPV-RPV,d)} = 881.3\times\lambda''_{j'12}\lambda''^{*}_{j'32}
\lambda''_{j12}\lambda''^{*}_{j32} \frac{(100GeV)^4}{m^2_{\tilde
u^{j'}_R} m^2_{\tilde u^j_R}}\;.\nonumber
\end{eqnarray}
We list various constraints on relevant RPV parameters and
corresponding values for $y$ in Table III.

For $y_{(SM-RPV)}$, we keep only the two terms proportional to
$\lambda_{cc1}$ and $\lambda_{uc1}$ since the other two terms are
proportional to $V_{ub}$. We obtain, $\lambda_{q\prime
q1}=A\lambda^2\lambda'_{i21}\times[\lambda'^*_{i13}-\lambda\lambda'^*_{i23}]$.
In Table III, $y_{(SM-RPV)_1}$ and $y_{(SM-RP)_2}$ indicate
contributions from the first and the second term in
$\lambda_{q\prime q1}$. Using constraints from \cite{Kundu:2004}, we
have $|\lambda'_{i21}\lambda'^*_{i13}|\lesssim 1.2\times{10^{-5}}$
and $|\lambda'_{i21}\lambda'^*_{i23}|\lesssim 5.0\times{10^{-5}}$,
each gives $y\approx 1\times 10^{-4}$. This value is much less than
the SM prediction.

For $y_{(RPV-RPV,\nu (l))}$, using the constraints
$|\lambda_{ij1}'^*\lambda_{i'j3}'|,~|\lambda_{i1j}'^*\lambda_{i'3j}'|
\lesssim  1.1\times 10^{-3}$, from \cite{Chemtob:2004} we find that
the corresponding upper bounds: $y_{(RPV-RPV,\nu)}\simeq 2.7\times
10^{-4}$ and $y_{(RPV-RPV,l)}\simeq 6.7\times 10^{-5}$.

As for the contribution $y_{(RPV-RPV,u)}$, there are four terms with
$j,j'$ take values $1$ or $2$. Taking the explicit constraints from
Ref.~\cite{Kundu:2004}, $|\lambda'_{i11}\lambda'_{i13}|\lesssim
8.0\times 10^{-4}$, $|\lambda'_{i11}\lambda'_{i23}|\lesssim
2.5\times 10^{-3}$, $|\lambda'_{i21}\lambda'_{i13}|\lesssim
1.2\times 10^{-5}$, and $|\lambda'_{i21}\lambda'_{i23}|\lesssim
5.0\times 10^{-5}$, we find that the dominant contribution is from
the case $j=j'=1$ which gives the upper bound $y_{(RPV-RPV,u)}\simeq
1.1\times 10^{-4}$. In the same way for the contribution
$y_{(RPV-RPV,d)}$, taking the constraints from
~\cite{Kundu:2004,Ghosh:2001mr}, the dominant part is from
$\lambda'_{ijj'}\lambda'^*_{i13}\lambda'_{i'31}\lambda'^*_{i'jj'}$
term with $j=j'=1$. We find the value for $y_{(RPV-RPV,d)}$ can be
as large as $7\times 10^{-3}$. This is about three times larger than
the SM contribution.

Contributions from $\lambda''$ are also constrained.
Ref.~\cite{Bar-Shalom:2002sv} considers the decay mode $B^-\to
\phi\pi^-$ and drives the upper bound
$\lambda''_{j21}\lambda''^*_{j32}< 6\times 10^{-5}$. Taking the same
bound for $|\lambda''_{j21}\lambda''^*_{j32}|_{j=2}$ and
$|\lambda''_{j21}\lambda''^*_{j32}|_{j=1,2}$ under the assumption
that there is no accidental cancellation, $y_{(SM-RPV)}$ and
$y_{(RPV-RPV,u(d))}$  are constrained to be small as can be seen
from Table III.

We conclude that if there is no accidental cancellations, for
$B_d^0-\bar B_d^0$ mixing, R-parity contribution to $y$ can be as
large as $7\times 10^{-3}$ which is about three times larger than
the SM prediction. This value is still difficult to be measured
experimentally. However, if there is accidental cancellation, $y$
could be bigger. Careful measurement of $y$ for $B^0_d -\bar B_d^0$
can provide valuable information about new physics beyond the SM.

\subsection{$B_s^0-\bar B_s^0$ mixing}
For $\lambda'$ contributions, we have
\begin{eqnarray}\label{eq:Bs1}
&&y_{(SM-RPV)}= -316.6\times\lambda_{q'q2}
\frac{(100GeV)^2}{m^2_{\tilde {e}^i_L}}\;,\nonumber\\
&&y_{(RPV-RPV,\nu)} = 77.7\times(100\mbox{GeV})^4\left\{
\frac{\lambda'_{j3i'}\lambda'^*_{j'2i'}}{m^2_{\tilde{d}^{i'}_R}}
\frac{\lambda'_{j'3i}\lambda'^*_{j2i}}{m^2_{\tilde{d}^i_R}}
\right. \nonumber\\
&&\hspace{2.4cm}+\left.\frac{\lambda'_{ji2}\lambda'^*_{j'i3}}{m^2_{\tilde{d}^i_L}}
\frac{\lambda'_{j'i'2}\lambda'^*_{ji'3}}{m^2_{\tilde{d}^{i'}_L}} +2
\frac{\lambda'_{j3i'}\lambda'^*_{j'2i'}}{m^2_{\tilde{d}^{i'}_R}}
\frac{\lambda'_{j'i2}\lambda'^*_{ji3}}{m^2_{\tilde{d}^i_L}}
\right\} \;,\nonumber\\
&&y_{(RPV-RPV,l)}=77.7\times\lambda'_{ji2}\lambda'^*_{j'i3}\lambda'_{j'i'2}\lambda'^*_{ji'3}
\frac{(100GeV)^4}{m^2_{\tilde{u}^i_L}m^2_{\tilde{u}^{i'}_L}}\;,\\
&&y_{(RPV-RPV,u)}=
233.1\times\lambda'_{ij2}\lambda'^{*}_{ij'3}\lambda'_{i'j'2}\lambda'^{*}_{i'j3}\frac{(100GeV)^4}{m^2_{\tilde
e^i_L}m^2_{\tilde e^{i'}_L}}  \;,\nonumber\\
&&y_{(RPV-RPV,d)} = 233.1 \times
\left(\lambda'_{i3j'}\lambda'^*_{i2j}\lambda'_{i'3j}\lambda'^*_{i'2j'}+
\lambda'_{ij2}\lambda'^*_{ij'3}\lambda'_{i'j'2}\lambda'^*_{i'j3}\right.\nonumber\\
&&\hspace{2.4cm}\left.-28\lambda'_{ijj'}
\lambda'^*_{i23}\lambda'_{i'32}\lambda'^*_{i'jj'} \right)
\frac{(100GeV)^4}{m^2_{\tilde {\nu}^i_L} m^2_{\tilde {\nu}^{i'}_L}}
\;.\nonumber
\end{eqnarray}
For $\lambda''$ contributions, we have
\begin{eqnarray}\label{eq:Bs2}
&&y_{(SM-RPV)} = -2.9\times\lambda_{221}''\lambda_{231}''^*\frac{(100GeV)^2} {m^2_{\tilde {d}^1_R} },\nonumber\\
&&y_{(RPV-RPV, u)}= 310.8\times\lambda''_{i12}\lambda''^{*}_{i'13}
\lambda''_{i'12}\lambda''^*_{i13}
\frac{(100GeV)^4}{m^2_{\tilde d^1_R} m^2_{\tilde d^{1}_R}}\;,\\
&&y_{(RPV-RPV,d)}= 1243\times\lambda''_{i'21}\lambda''^{*}_{i'31}
\lambda''_{i21}\lambda''^{*}_{i31} \frac{(100GeV)^4}{m^2_{\tilde
u^{i'}_R} m^2_{\tilde u^i_R}}\;.\nonumber
\end{eqnarray}

\begin{table}[thb!]
\begin{center}
{\small
\begin{tabular}{ccccccccr}
\hline\hline
 \ RPV parameters  \ & \ Bounds [Processes] \ & \ Estimate \  & \ Our bounds on RPV  \
\\ \hline
$|\lambda'_{i23}\lambda'^*_{i22}|$\ &  $8.2\times{10^{-3}}\;[\bar
B_sB_s]$ \ & $y_{(SM-RPV)}\simeq 0.11$
  \ & $7.4\times 10^{-3}\, (7.8\times 10^{-4})$ \\
  $|\lambda'_{j3i'}\lambda'^*_{j'2i'}|$, $|\lambda_{ij2}'^*\lambda_{i'j3}'|$ \ & $1.5\times 10^{-3}$~$[B\to X_s\nu\bar\nu]$
  & $y_{(RPV-RPV,\nu)}\simeq 7\times 10^{-4}$ & - \\
  $|\lambda_{ij2}'^*\lambda_{i'j3}'|$ \ & $1.5\times 10^{-3}$~$[B\to X_s\nu\bar\nu]$
  & $y_{(RPV-RPV,l)}\simeq 1.7\times 10^{-4}$ & -\\
  $|\lambda'_{ij2}\lambda'^{*}_{ij'3}|_{j,j'\neq3}$ \ & $5.16\times 10^{-2}$ $[B_s\bar{B_s}]$ & $y_{(RPV-RPV,u)}\simeq 0.06$ &
  $2.0\times{10^{-2}}\,(6.6\times 10^{-3})$ \\
  $|\lambda_{ijj'}'\lambda_{i23}'^{*}\lambda_{i'32}'\lambda_{i'jj'}'^{*}|_{j,j'\neq3}$ \ & see text & $y_{(RPV-RPV,d),3}
  \simeq 0.26$ & $1.5\times 10^{-5}\,(1.6\times 10^{-6})$ \\
  $|\lambda_{221}''\lambda_{231}''^{*}|$ \ & $1.01\times{10^{-2}}$ [$B\to \bar{K}\pi$]  & $y_{(SM-RPV)}\simeq 2.9\times 10^{-2}$ &
   $3.3\times{10^{-2}}\, (3.5\times 10^{-3})$ \\
  $|\lambda_{i12}''\lambda_{i13}''^*|_{i\neq3}$  \ & see text & $y_{(RPV-RPV, u)}$, see text & $1.77\times
10^{-2}\,(5.7\times 10^{-3})$ \\
  $|\lambda''_{j'21}\lambda''^{*}_{j'31}|$ \ & $1.2\times{10^{-3}}$ [$B^+\to \pi^+K^0$] & $y_{(RPV-RPV,d)}\simeq
  1.8\times 10^{-3}
  $ & - \\
  \hline\hline
\end{tabular}}
\caption{Upper limits on parameters from
\cite{Barbier:2004ez,Allanach:1999ic,Chemtob:2004,Kundu:2004,Nandi:2006,Ghosh:2001mr,Chakraverty:2000df}
and corresponding values for $y$. The numbers in the brackets
correspond to the case, when central values for $y_{SM}$ and
$y_{Exp.}$ are used to put the constraints. For each number see the
text for the explanation.}
\end{center}
\end{table}

We list the constraints on the RPV parameters from
\cite{Barbier:2004ez,Chemtob:2004} and the corresponding values for
the mixing parameter $y$ in Table IV.

There are several terms contributing to $y$ from $\lambda'$. For
$y_{(SM-RPV)}$ case we again drop terms proportional to $V_{ub}$,
and have, $\lambda_{q\prime
q2}=A\lambda^2\lambda'_{i22}\times[\lambda'^*_{i23}+\lambda\lambda'^*_{i13}]$.
We are using constraints from Ref.~\cite{Nandi:2006} we have
$|\lambda'_{i23}\lambda'^*_{i22}|\lesssim 8.2 \times{10^{-3}}$ and
from Ref.~\cite{Ghosh:2001mr}
$|\lambda'_{i13}\lambda'^*_{i22}|\lesssim 2.48 \times{10^{-3}}$. The
first term dominates and gives $y_{(SM-RPV)}\simeq 0.1$, which is of
order of SM prediction $y_{SM}\simeq 0.078$ and may have measurable
effect.

For $y_{(RPV-RPV,\nu)}$, we have three contributions. For first and
second contributions using the following conditions on RPV
parameters $|\lambda'_{j3i'}\lambda'^*_{j'2i'}|$,
$|\lambda_{ij2}'^*\lambda_{i'j3}'|$ $\lesssim1.5\times 10^{-3}$
\cite{Barbier:2004ez}, we get $y_{(RPV-RPV,\nu)}\simeq 1.7\times
10^{-4}$. For the last term, we obtain $y_{(RPV-RPV,\nu)}\simeq
3.5\times 10^{-4}$. If we simply add them together we will get
$y_{(RPV-RPV,\nu)}\simeq 7\times 10^{-4}$.

For $y_{(RPV-RPV,l)}$, the situation is the same as the second term
of ${(RPV-RPV,\nu)}$ case.

In the case for $y_{(RPV-RPV,u)}$, if one uses the individual
constraints from \cite{Chemtob:2004,Nandi:2006,Ghosh:2001mr}
$(\lambda_{i12}'\lambda_{i13}'^{*},\lambda_{i22}'\lambda_{i23}'^{*},
\lambda_{i12}'\lambda_{i23}'^{*},\lambda_{i22}'\lambda_{i13}'^{*})=(1.63\times
10^{-3},8.2\times 10^{-3},5.16\times 10^{-2},2.48\times 10^{-3})$
can get for each contribution $y_{(RPV-RPV,u)}\simeq(6.2\times
10^{-4},1.6\times 10^{-2},6.0\times 10^{-2})$. If we keep the
dominant interference term  we will get $y\simeq 0.06$.

For $y_{(RPV-RPV,d)}$ case we have three contributions. The
contribution of the first term into $y$ is small, about
$1.6\times{10^{-3}}$. The dominant contributions here are coming
from squares of $|\lambda'_{i31}\lambda'^{*}_{i21}|\simeq 1.29\times
10^{-3}$ $[B^-\to K^-\pi_0]$ \cite{Ghosh:2001mr} and
$|\lambda'_{i32}\lambda'^{*}_{i22}|\simeq 2.3\times 10^{-3}$
$[B^0\to MM]$ \cite{Chemtob:2004}. The second term is just the same
as in $y_{(RPV-RPV,u)}$ case, considered above. So here we have
$y_{(RPV-RPV,l)}\simeq 0.06$. The last term is enhanced with the
large coefficient. Here the dominant contributions are coming from
$|\lambda'_{i22}\lambda'^{*}_{i23}\lambda'_{i'32}\lambda'^{*}_{i'22}|\simeq
18.9\times 10^{-6}$ and
$|\lambda'_{i12}\lambda'^{*}_{i23}\lambda'_{i'32}\lambda'^{*}_{i'12}|\simeq
20.6\times 10^{-6}$
\cite{Nandi:2006,Ghosh:2001mr,Chemtob:2004,Kundu:2004}. If we simply
add them together, their contribution will be $y_{(RPV-RPV,d)}\simeq
0.26$. So, here we can conclude, that in $y_{(RPV-RPV,d)}$ case one
also can expect large effects for $y$.

For $\lambda''$ contribution to $y_{(SM-RPV)}$, we have
$|\lambda''_{221}\lambda''^{*}_{231}|\lesssim 1.01\times 10^{-2}$
from $B\to \bar{K}\pi$ \cite{Ghosh:2001mr}, which gives
$y_{(SM-RPV)}\simeq 2.9\times 10^{-2}$.

For the $y_{(RPV-RPV, u)}$ case no direct constraint on
$|\lambda_{i12}''\lambda_{i13}''^*|_{i=1,2}$ exists. However, if one
assumes that
$|\lambda_{i12}''\lambda_{i13}''^*|_{i=1,2}\approx|\lambda_{i12}''\lambda_{i13}''^*|\lesssim
1.2\times 10^{-3}$ \cite{Chakraverty:2000df}, $y_{(RPV-RPV,
u)}\simeq 4.5\times 10^{-4}$.

For the last case of Eq.~(\ref{eq:Bs2}) from
\cite{Chakraverty:2000df} we have
$|\lambda''_{j'21}\lambda''^{*}_{j'31}|\lesssim1.2\times{10^{-3}}$,
which gives $y_{(RPV-RPV,d)} \approx 1.8\times 10^{-3}$.

We note that present constraints on the RPV parameters still allow
large $y_{(SM-RPV)}$, $y_{(RPV-RPV,u)}$ and $y_{(RPV-RPV,u)}$ from
$\lambda'$ interaction. One can turn the argument around to
constrain the relevant RPV parameters by requiring that the new
contributions do not exceed the allowed range for the difference of
SM prediction and experimental values. We have carried out an
analysis, taking the one sigma range values $y_{SM}=[0.054,0.101]$
and $y_{Exp.}=[0.022,0.151]$, and assumed constructive contributions
between the SM and new contributions, to obtain the bounds for each
individual terms. Similar analysis has been performed for
$\lambda''$ cases. The bounds are listed in Table IV in the last
column. These bounds are new ones.

\section{Conclusion}
In this paper we have explored the influence of SUSY R-parity
violation contributions for the lifetime difference $y$ on the $D^0
-\bar{D}^0$ and $B^0_{d,s} - \bar{B}^0_{d,s}$ systems. We have
obtained general expressions for new physics contributions to $y$
from effective four fermion operators including SM-NP interference
and pure new physcis contributions. We find that in general R-parity
violating contribution to $D^0 - \bar D^0$ mixing, and $B_{d}^0 -
\bar B_{d}^0$ to be small. There may be sizable contribution to
$B_s^0 -\bar B_s^0$ mixing. We also obtain some interesting bounds
on R-parity violating parameters using known Standard Model
predictions and experimental data.

\newpage
\noindent {\bf Acknowledgments}$\,$ This work was supported in
part by the NSC and NCTS.

\end{document}